\documentclass[%
 reprint,
 amsmath,amssymb,
 aps,
 nolongbibliography
]{revtex4-2}

\usepackage{graphicx}
\usepackage{dcolumn}
\usepackage{bm}
\usepackage[dvipsnames]{xcolor} 

\usepackage{ulem}
\usepackage{xcolor}

\usepackage{soul}  
\usepackage{hyperref}

\begin{document}

\title{Collective behaviors of self-propelled particles with {tunable alignment angles}}
\author{Zichen Qin}
\author{Nariya Uchida}
 \email{uchida@cmpt.phys.tohoku.ac.jp}
\affiliation{
Department of Physics, Tohoku University, Sendai 980-8578, Japan
}
\date{\today}
\begin{abstract}
We present a novel aligning active matter model by extending the nematic alignment rule in self-propelled rods 
to tunable alignment angles, as represented by collision of cone-shaped particles.
Non-vanishing alignment angles introduce frustration in the many-body interactions, 
and we investigate its effect on the collective behavior of the system. 
Through numerical simulations of an agent-based microscopic model, we found that the system exhibits distinct phenomenology compared to the original self-propelled rods. In particular, anti-parallel bands are observed in an intermediate parameter range. The linear stability analysis of the continuum description derived from the Boltzmann approach demonstrates qualitative consistency with the microscopic model,
while frustration due to many-body interactions
in the latter destabilizes homogeneous nematic order
over a wide range of the alignment angle.
\end{abstract}


\maketitle

\section{\label{sec:level1}Introduction}

Aligning self-propelled particles represent a typical active matter system, where active particles consume energy to generate directed motion and tend to align their velocities, leading to the formation of flocks~\cite{vicsek2012collective,chate2020dry,Bechinger2016active}. 
This framework has long been used to study collective motion in various natural systems, 
such as groups of animals~\cite{reynolds1987flocks,couzin2002collective} 
(including fish~\cite{aoki1982simulation,huth1992simulation} and birds~\cite{cavagna2014bird}),
bacterial colonies~\cite{peruani2012collective}, and cytoskeletal filaments~\cite{kruse2004asters,schaller2010polar}.
A prototypical model of aligning self-propelled particles is the Vicsek model~\cite{vicsek1995novel}, 
where nearby particles experience polar interactions that tend to align their directions in parallel 
and exhibit collective behaviors distinct from their passive counterpart, such as moving bands~\cite{chate2008collective}.
It has been extended to four classes of models by the symmetry of the self-propelling motion and alignment rules.
Apart from the original Vicsek model, there are
dry active nematics with apolar motion and nematic alignment~\cite{chate2006simple},
self-propelled rods with polar motion and  nematic alignment~\cite{ginelli2010large}, 
and Vicsek-shake model with apolar motion and polar alignment~\cite{mahault2018self}.
{Here, apolar motion means that reversals in the self-propulsion direction occur at a certain rate.}
Among them, self-propelled rods 
{have} 
garnered extensive attention~\cite{bar2020self}. 
In the original numerical study~\cite{ginelli2010large}, it was demonstrated that the system exhibits homogeneous nematic order 
under low-noise conditions and 
transitions to a homogeneous disordered state at high noise levels. 
At intermediate noise levels, phase separation occurs, characterized by high-density nematic bands surrounded by a disordered gas. 
Subsequent experimental studies revealed that the self-propelled rods model can be adapted to describe 
many biological systems, such as bacteria~\cite{peruani2012collective,nishiguchi2017long}, microtubules~\cite{sumino2012large,afroze2021monopolar}, and 
{actin filaments}~\cite{huber2018emergence}. 
Microtubules in gliding assays exhibit counter-clock-wise rotation due to the chirality of the filament, which leads 
to formation of large-scale vortices~\cite{sumino2012large} or polar flocking~\cite{afroze2021monopolar}.
An analysis of binary collision data in actomyosin motility assays~\cite{huber2018emergence}, 
{supported by numerical simulations using an active polymer model,} 
suggested that 
introducing a tunable polar bias to the alignment rule can significantly improve the model's realism. 
Theoretical analysis based on continuum models~\cite{denk2020pattern} also indicates  
{that a polar bias} in the alignment rule
leads to coexistence of polar and nematic patterns, thereby reproducing the experimental observations.

{However, the alignment rules considered in discrete Vicsek-type models have been
restricted to parallel and anti-parallel alignment,
while non-vanishing alignment angles other than 0 or $\pi$ could enrich the collective behavior.
}
{
As a related study,
Refs.~\cite{kruk2018self,kruk2020traveling} introduced a phase lag 
in a Kuramoto-Vicsek model that merges synchronization and self-propulsion~\cite{degond2014hydrodynamics,liebchen2017collective}. 
The phase lag induces rotation of synchronized clusters,  
accompanied by a coexisting disordered state for a phase lag close to $\pi/2$.}
Refs.~\cite{moran2022particle,moran2022shape} 
investigated an active colloidal particle system with shape anisotropy, 
revealing {the effects of particle shape on motility-induced phase separation.}
Specifically, they studied triangular active colloidal particles with a vertex-forward self-propulsion direction, which attracted our attention. 
{Furthermore, Refs.~\cite{kumar2014flocking} investigated tapered rods contacting with an underlying vibrated surface and interacting through a medium of spherical beads, whose particle shape is similar to that considered in the present study.}

\begin{figure}[tb]
\includegraphics[width=0.45\textwidth]{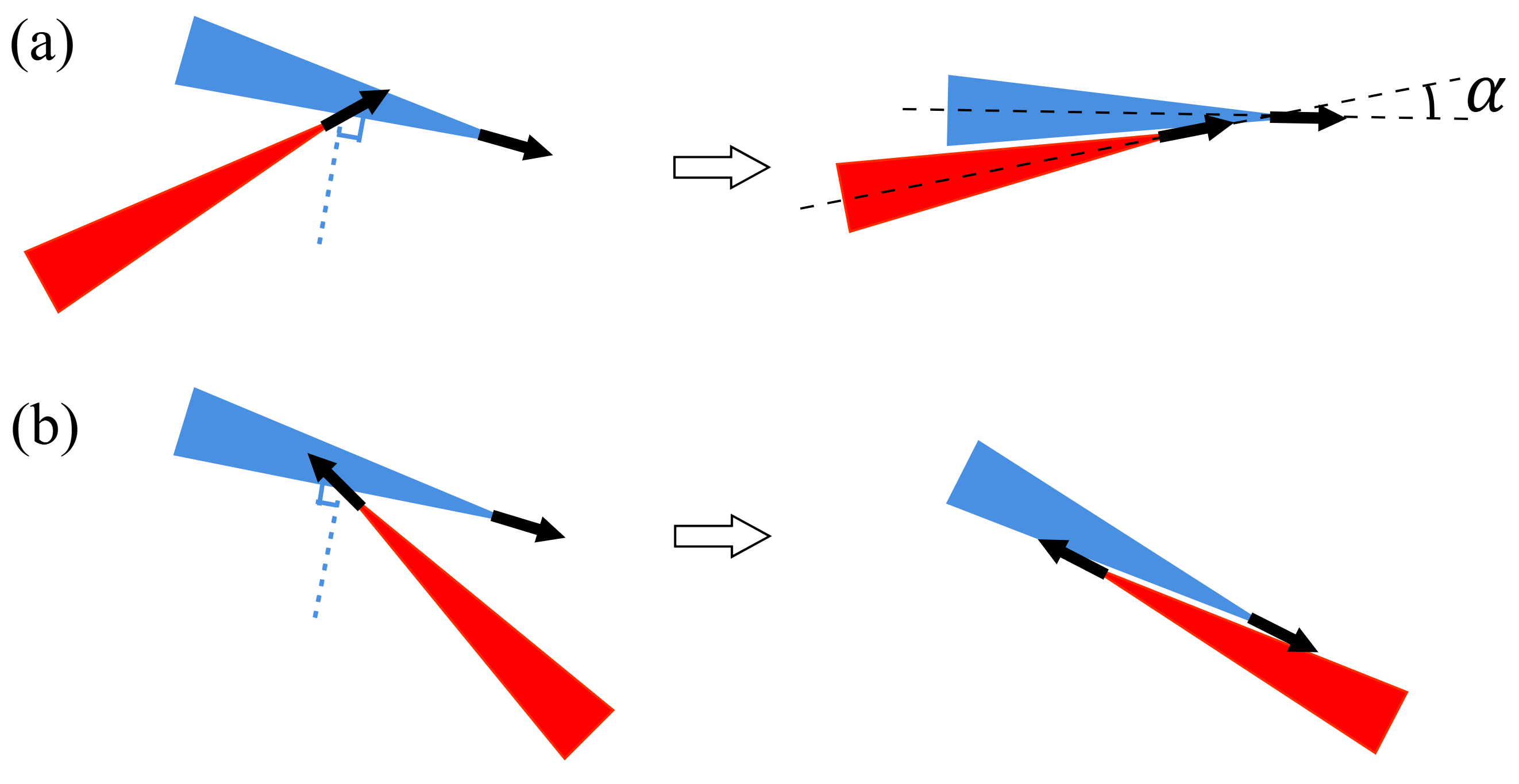}
\caption{\label{fig1} Illustration of the alignment {rule for} cone-shaped particles characterized by the apex angle $\alpha$. (a) A collision at a small angle (with the heading angle difference smaller than $(\alpha + \pi)/2$) causes the angle difference $\alpha$. (b) A collision at a large angle results in the anti-parallel alignment.}
\end{figure}

\begin{figure*}[tb]
\includegraphics[width=\textwidth]{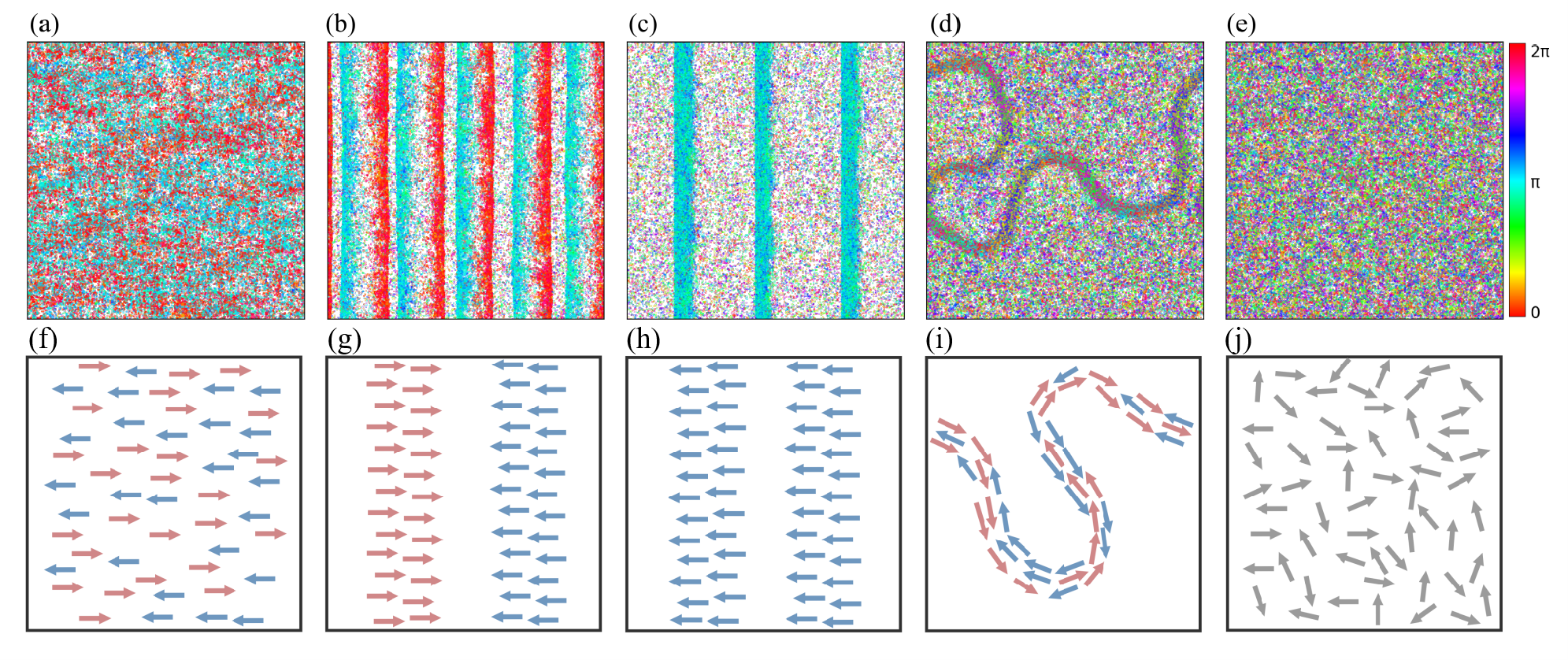}
\caption{\label{fig2} Typical snapshots of {collective} states at different values of $\alpha=0.05$. (a) $\alpha=0.05$: homogeneous nematic ordered state. (b) $\alpha=0.35$: anti-parallel polar bands. (c) $\alpha=0.66$: parallel polar bands. (d) $\alpha=0.74$: chaotic nematic bands. (e) $\alpha=0.75$: homogeneous disordered state. The particle orientation is represented by color.
{(f)-(j) Schematic illustrations of the different states corresponding to (a)-(e).
Note that for the anti-parallel polar bands, the red and blue bands move in opposite directions and penetrate each other.}
}
\end{figure*}

In this study, we consider {cone-shaped} self-propelled particles constrained to move on a two-dimensional surface. Thanks to recent advancements in synthesizing active particles with anisotropic shapes~\cite{wang2022engineering}, this system holds promise for experimental investigation. 
{For simplicity,  we represent each particle as a triangular shape {of the apex angle $\alpha$},
as illustrated in Fig. 1.
} 
When the incoming angle difference is small, two particles align with 
the angular difference of $\alpha$. When the angle difference exceeds a threshold, the particles align in opposite directions, consistent with the original nematic alignment rule. This threshold is determined by the direction perpendicular to the side of the particle. For $\alpha=0$, the model reduces to the original self-propelled rods model. For non-vanishing $\alpha$, 
a collision between three or more particles induces frustration. 
Therefore, $\alpha$ serves as a key parameter controlling frustration in the system.

We perform numerical simulations based on an agent-based microscopic model and identify distinct phenomenological differences compared to the original {self-propelled rods} model. In Section II, we define the model and present the simulation results. 
Notably, in the intermediate parameter range, we observe polar bands and 
metastable chaotic nematic bands. In Section III, 
we develop
a continuum description based on the Boltzmann approach
and conduct a linear stability analysis of the homogeneous solution. The results qualitatively capture the macroscopic behavior observed in the microscopic model,
{although we find quantitative differences due to many-body interactions neglected in the Boltzmann approach}.
We discuss the role of frustration and compare our results with those of related studies in Section IV.

\section{\label{sec:level1}Microscopic Model}

{In our model,} $N$ particles move within a two-dimensional square domain of size $L \times L$ with periodic boundary conditions. 
The orientation and position of particle $j$ are updated at discrete time steps as follows:
\begin{align}
\theta _{j}^{t+1} &=\mathrm{arg}\left[\sum _{k\sim j} g\left( \theta _{j}^{t} ,\theta _{k}^{t}\right) e^{i\theta _{k}^{t}}\right] +\eta \xi _{j}^{t} 
\\
\mathbf{r}_{j}^{t+1} &=\mathbf{r}_{j}^{t} + v_{0} (\cos \theta_{j}^{t+1}, \sin \theta_{j}^{t+1})
\label{eq1}
\end{align}
The summation extends over all neighboring particles $i$ within a unit distance. $\xi$ represents a noise term uniformly distributed in the interval $[-\frac{\pi}{2}, \frac{\pi}{2}]$, while $\eta$ denotes the noise intensity. The alignment rule is characterized by the function $g(\theta_1, \theta_2)$, which is defined as follows:
\begin{equation}
g(\theta _{1} ,\theta _{2})=\left\{
\begin{aligned}
&e^{i\ \mathrm{sign}[\Delta(\theta_1,\theta_2)]\alpha} &\mathrm{if}\  |\Delta(\theta_1,\theta_2)|<\frac{\alpha+\pi}{2} \\
&-1 &\mathrm{if}\ |\Delta(\theta_1,\theta_2)|>\frac{\alpha+\pi}{2}
\end{aligned}
\right.\label{eq3}
\end{equation}
in which $\Delta(\theta_1,\theta_2)$ is the angular difference $\theta_1-\theta_2$ with the result wrapped on the interval $[-\pi,\pi]$.

\begin{figure}[b]
\includegraphics[width=0.455\textwidth]{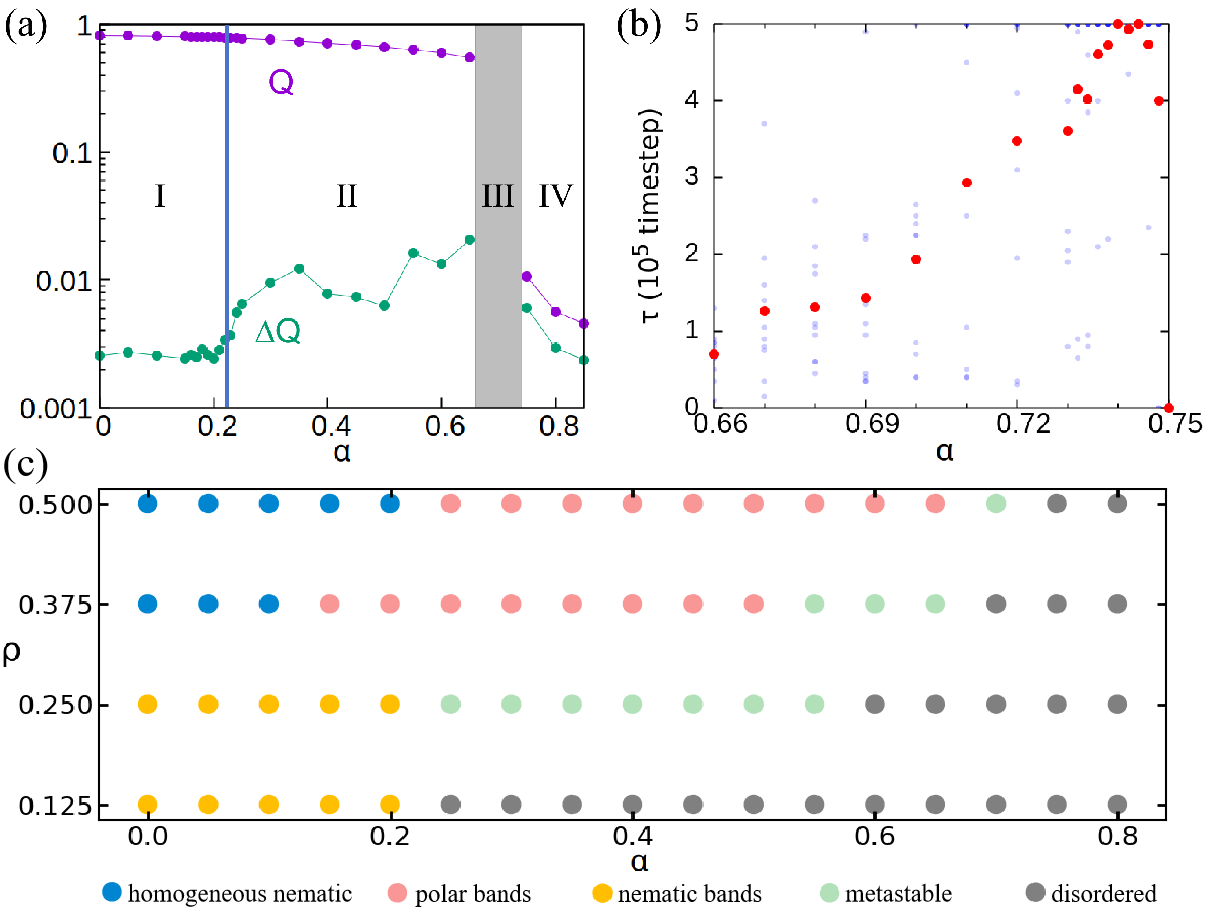}
\caption{\label{fig3} 
(a) Variation of the global nematic order parameter $Q$ and its root mean square (rms) fluctuation $\Delta Q$ with $\alpha$. The average density is fixed at $\bar{\rho} = N/L^2 = 0.5$.
Region I: homogeneous nematic state. Region II: polar bands {(either anti-parallel or parallel).} 
Region III: metastable chaotic nematic bands. 
Region IV: homogeneous disordered state {where $Q$ is close to zero.}
In region III, instead of plotting the order parameter, 
we present the average lifetime of the chaotic nematic band state in (b). (b) Average lifetime $\tau$ of the chaotic nematic bands. The maximum timestep of each simulation is $5 \times 10^5$. The red dot represents the average taken over 10 samples, while the blue dots show all data from these 10 samples.
(c) Numerical simulation results at other densities.
}
\end{figure}

\begin{figure*}
\includegraphics[width=1\textwidth]{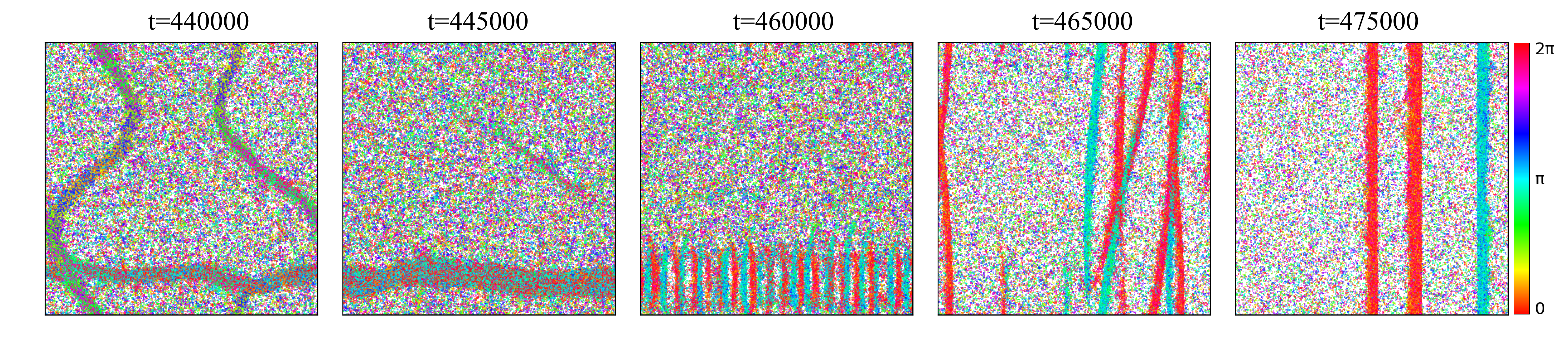}
\caption{\label{fig4} 
Snapshots of the time evolution in region III are shown, illustrating the transition from chaotic nematic bands to anti-parallel polar bands. In this simulation $\alpha = 0.73$. The timestep is indicated above each snapshot, and the particle orientation is represented by color.
}
\end{figure*}

In this study, we fix the noise strength at $\eta = 0.2$ and the system size at $L = 512$. We use a homogeneous nematic-ordered initial condition to establish nematic order parallel to the boundaries of the simulation box, thereby minimizing potential finite-size effects caused by non-parallel bands. By varying $\alpha$, we observe distinct {collective} states as illustrated in Fig.~\ref{fig2}. 
Due to the geometric meaning of $\alpha$ {as the apex angle of the particle,} 
its value should be restricted to an appropriate range. {We consider the range $0 \le \alpha \le 0.85$ in our simulations.} 
When $\alpha$ is either sufficiently small or large, the system exhibits a homogeneous nematic-ordered state and a homogeneous disordered state, respectively. At small $\alpha$, the frustration effect is weak, 
{and} the system maintains the same homogeneous nematic-ordered state 
as in the original {self-propelled rods} model  {[Fig.~\ref{fig2}(a)]}. 
Conversely, at large $\alpha$, where the frustration effect is pronounced, the system fails to achieve synchronization through alignment interactions, resulting in a homogeneous disordered state {[Fig.~\ref{fig2}(e)]}. 
In the intermediate parameter range, we observe the emergence of 
polar bands, {which align either anti-parallel [Fig.~\ref{fig2}(b)]
or parallel [Fig.~\ref{fig2}(c)] to each other},
and chaotic nematic bands {[Fig.~\ref{fig2}(d)]}. 
It should be noted that Fig.~\ref{fig2} (a), (b), and (c) show the final states of the system
{in our simulations}, 
while {the chaotic bands in} (d) represents a 
{transient} state
that {transitions}  to polar bands 
{after a long time,}
which will be discussed in detail later. 
{Furthermore,} the parallel polar bands shown in (c) is preceded by chaotic nematic bands.

In Fig.~\ref{fig3}(a), we plot the global nematic order parameter $Q$ and its root mean square fluctuation $\Delta Q$ versus $\alpha$. Different collective states are observed in the regions I-IV. The region I ($0 \le \alpha < 0.23$) corresponds to the homogeneous nematic state with  large $Q$ and small $\Delta Q$ values. 
The polar bands are observed in region II ($0.23 \le \alpha < 0.66$). 
{Besides the parallel polar bands seen in the Vicsek model, which move in the same direction and exhibit global polar order, the anti-parallel case is also observed, in which the polar bands propagate in opposite directions.} As a result, they exhibit local polar order but maintain global nematic order. 
{Within this parameter range, the system randomly enters either of these two states due to differences in initial conditions and noise.} 
The chaotic nematic bands represent a metastable state, corresponding to the parameter region III ($0.66 \le \alpha < 0.75$). Here, metastability means that the system initially forms chaotic nematic bands that continuously bend and deform for an extended period of time; however, at some point during the simulation, these bands transition into polar bands which move in either anti-parallel or parallel directions. The duration of the chaotic band state and the final band configuration (parallel or anti-parallel) depends on the initial conditions and noise samples. Fig.~\ref{fig3}(b) shows the lifetime $\tau$ of the chaotic nematic bands. When $\alpha$ is below 0.71, most simulation samples exhibit a short lifetime, indicating that the chaotic nematic bands are unstable and quickly transition to polar bands. As $\alpha$ increases, particularly near 0.74, the chaotic nematic bands in most samples reach the maximum simulation time without transitioning, demonstrating increased stability. However, at $\alpha = 0.75$, the chaotic nematic bands disappear entirely, and the system exhibits a homogeneous disordered state.

The typical transition process from chaotic nematic bands to polar bands is illustrated in Fig.~\ref{fig4}. Initially, the chaotic nematic bands evolve into a straight nematic band structure. However, rather than bending and deforming to resume their previous chaotic behavior, the nematic band develops local instabilities, which give rise to the formation of narrow anti-parallel polar bands. Over time, the width of these polar bands gradually increases, ultimately leading to a global pattern transition. 
{It should be noted that the lifetime measured here is that of the nematic state, representing the first metastable state entered from the initial condition. However, in the region near the transition to disorder, it was also observed that the system first enters a disordered state from the initial condition, subsequently transitioning to a nematic band after a certain period.}

{Fig.~\ref{fig3} (c) illustrates the system's behavior with changing $\alpha$ at other densities. At $\rho = 0.375$, the same four regions as observed at $\rho = 0.5$ are present. However, when the density is 0.25, the system exhibits a qualitative change in its behavior. As $\alpha$ increases, the system first enters a nematic band state. With a further increase in $\alpha$, the nematic band becomes unstable and transitions to a metastable state after certain time steps, consistent with previous observations. Upon further decreasing the density to 0.125, the system transitions directly from the nematic band state to a disordered state, and the metastable region disappears.}

\section{\label{sec:level1}Boltzmann Approach}
{Following} previous studies \cite{bertin2006boltzmann,bertin2009hydrodynamic}, 
{we formulate a continuous and coarse-grained description 
of our microscopic model} 
through the Boltzmann equation 
\begin{equation}
    \partial_t f(\mathbf{r},\theta,t)+v_0\mathbf{e}(\theta)\cdot\nabla f(\mathbf{r},\theta,t)=I_{\mathrm{dif}}[f]+I_{\mathrm{col}}[f] ,
    \label{eq4}
\end{equation}
which governs the time evolution of the one-particle distribution function $f(\mathbf{r},\theta,t)$. 
{Each particle changes its orientation due to self-diffusion and binary-collision events.
The self-diffusion events occur with probability $\lambda$ per unit time. 
The collision frequency is given by the kernel $K( \phi )=4v_{0} r_{0} |\sin\frac{\phi }{2} |$, 
where $v_{0}$ denotes the self-propulsion speed, $r_{0}$ is the interaction radius of a particle, 
and $\phi$ is the angular difference between the two particles.
The resultant change in the orientational distribution is described by the self-diffusion integral}
\begin{align}
I_{\mathrm{dif}} =
&-\lambda f(\boldsymbol{r} ,\theta ,t)
\nonumber\\ 
&\hspace{-5mm} 
+\lambda \int _{-\pi }^{\pi } d\theta '
\int _{-\infty }^{\infty } d\eta \ P_{\sigma }( \eta ) \ 
\delta _{2\pi }( \theta '+\eta -\theta ) \ f(\boldsymbol{r} ,\theta ',t),
\label{eqa1}
\end{align}
{and the collision integral} 
\begin{align}
I_{\mathrm{col}} =
&-f(\boldsymbol{r} ,\theta ,t)
\int _{-\pi }^{\pi } d\theta '\ K( \theta '-\theta ) \ f(\boldsymbol{r} ,\theta ',t)  
\nonumber\\
& +\int _{-\pi }^{\pi } d\theta _{1}\int _{-\pi }^{\pi } d\theta _{2}\int _{-\infty }^{\infty } d\eta \ P_{\sigma }( \eta ) \ K( \theta _{2} -\theta _{1}) 
\nonumber\\
& \cdot f(\boldsymbol{r} ,\theta _{1} ,t) \ f(\boldsymbol{r} ,\theta _{2} ,t) \ 
\delta _{2\pi }( \Psi ( \theta _{1} ,\theta _{2}) +\eta -\theta),
\label{eqa2}
\end{align}
in which $P_{\sigma }( \eta )$ is the distribution of a Gaussian noise, and $\delta_{2\pi}(x) = \sum_{n=-\infty}^{\infty} \delta(x-2n\pi)$ is the periodically extended Dirac's $\delta$-function. The constant speed $v_0$, interaction radius $r_0$ and also $\lambda$ are effectively set to unity by rescaling. 
{The interaction kernel $\Psi(\theta_1,\theta_2)$ depends only on 
the angle difference $\phi =\theta_2 -\theta_1$, and}
$\Psi (0 ,\phi)=H(\phi)$ encodes the alignment rule shown in Fig.~\ref{fig1}:
\begin{equation}
\begin{aligned}
H(\phi)=\left\{
\begin{aligned}
&\frac{\phi }{2} +\frac{\pi }{2}  ,\ \ \ \left( -\pi ,-\frac{\alpha+\pi}{2}\right)\\
&\frac{\phi }{2} +\frac{\alpha}{2}, \  \ \left( -\frac{\alpha+\pi}{2} ,\ 0\right)\\
&\frac{\phi }{2} -\frac{\alpha}{2}, \ \ \left( \ 0,\ \ \frac{\alpha+\pi}{2}\right)\\
&\frac{\phi }{2} -\frac{\pi }{2}, \ \ \ \left( \ \frac{\alpha+\pi}{2} ,\ \pi \right)
\end{aligned}
\right.
\end{aligned}
\label{eqa3}
\end{equation}
By expanding $f(\mathbf{r},\theta,t)$ in Fourier series of its angular variable $\theta$, Eq.~(\ref{eq4}) can be rewritten as the infinite hierarchy of equations
\begin{equation}
\begin{aligned}
\partial _{t} f_{k} +\frac{1}{2}\left( \nabla f_{k-1} +\nabla ^{*} f_{k+1}\right) =( P_{k} -1) f_{k} \\ +\sum _{q=-\infty }^{\infty }( P_{k} I_{k,q} -I_{0,q}) f_{q} f_{k-q} ,
\end{aligned}
    \label{eq5}
\end{equation}
{where 
$f_{k}(\boldsymbol{r} ,t) =\int _{-\pi }^{\pi } d\theta \ f(\boldsymbol{r} ,\theta ,t) \ e^{ik\theta}$ $(k=1,2,3,\ldots)$
are the Fourier coefficients.
{ For $k=0$, this corresponds to the density field
$\rho(\boldsymbol{r} ,t) =\int _{-\pi }^{\pi } d\theta \ f(\boldsymbol{r} ,\theta ,t) $.
}
Here, $P_{k} =e^{-\sigma ^{2} k^{2} /2}$ 
and
$I_{k,q} =\frac{1}{2\pi }\int _{-\pi }^{\pi } d\phi\ K( \phi ) e^{-iq\phi +ikH( \phi )}$ are introduced.
}

\begin{figure}[]
\includegraphics[width=0.45\textwidth]{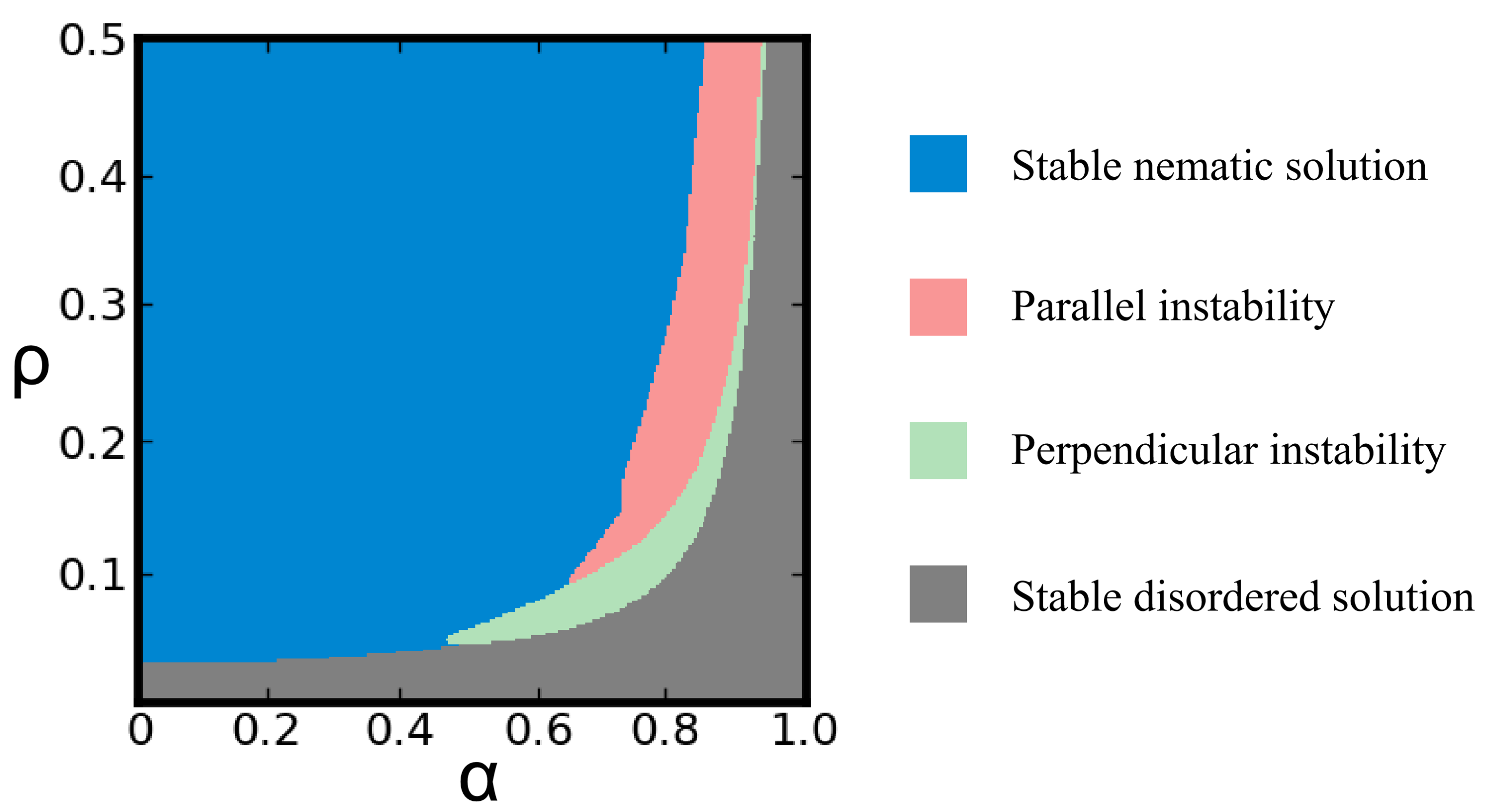}
\caption{\label{fig5} 
Stability diagram in the $\alpha$-$\rho$ parameter space with $\sigma=0.1$ fixed. The linear stability analysis of the homogeneous solution reveals four distinct regions: (1) stable nematic solutions, (2) unstable nematic solutions with the most unstable wavevector parallel to the nematic order (parallel instability), (3) unstable nematic solutions with the most unstable wavevector perpendicular to the nematic order (perpendicular instability), and (4) stable disordered solutions.
}
\end{figure}

Previous studies have revealed the fact that modes with sufficiently large $|k|$ can be neglected even beyond the ordering transition. Furthermore, the results derived under this assumption closely align with experimental observations and numerical simulations based on the Boltzmann equation \cite{denk2020pattern,denk2016active}. Following this method, we truncate the infinite hierarchy of equations at $k_{\mathrm{max}}=16$ and conduct linear stability analyses. By setting all spatial and temporal derivatives to zero, the
stationary homogeneous version of Eq.~(\ref{eq5}) is obtained as
\begin{equation}
\begin{aligned}
0=( P_{k} -1) |f_{k} |+\sum _{q=-\infty }^{\infty } J_{k,q} |f_{q} ||f_{k-q} | ,
\end{aligned}
\label{eq6}
\end{equation}
in which $J_{k,q}$ are defined as $J_{k,q} =P_{k} I_{k,q} -I_{0,q}$. Neglecting all Fourier modes with $k>k_{\mathrm{max}}$, we numerically solve the remaining equations in Eq.~(\ref{eq6}) and get the spatially homogeneous solution $f_k^{(0)}$. In the gray region of Fig.~\ref{fig5}, only homogeneous disordered solutions were found, characterized by $f_k^{(0)} = 0$ for $1 \leq k \leq k_{\text{max}}$. In contrast, in the other regions, we obtained homogeneous nematic-ordered solutions, where all remaining odd modes vanish while even modes are positive.

We then studied the linear stability of these solutions. By introducing small 
perturbations {as} $\rho=\bar\rho+\delta\rho{(\boldsymbol{r}, t)}$ and 
$f_k=f_k^{(0)}+\delta f_k{(\boldsymbol{r}, t)}$, 
{we can express Eq.~(\ref{eq5}) as}
\begin{equation}
\begin{aligned}
\partial _{t} \delta f_{k} =-\frac{1}{2}\left( \nabla \delta f_{k-1} +\nabla ^{*} \delta f_{k+1}\right) +( P_{k} -1) \delta f_{k} \\+\sum _{q=-\infty }^{\infty }( J_{k,q} +J_{k,k-q}) |f_{k-q}^{(0)} |\delta f_{q}.
\end{aligned}
\label{eq7}
\end{equation}
Considering that the perturbations take wave-like form $\delta\rho=\delta\rho_\mathbf{q}e^{i\mathbf{q}\cdot \boldsymbol{r}+\Gamma_{\mathbf{q}}t}$ and $\delta f_k=\delta f_{k,\mathbf{q}}e^{i\mathbf{q}\cdot \boldsymbol{r}+\Gamma_{\mathbf{q}}t}$, 
we solve the linear set of equations for the maximal real part of 
eigenvalue $\Gamma_{\mathbf{q}}$ in the wave vector space. 
A positive maximum value indicates the instability of corresponding homogeneous solution.

The results of this calculation are presented in Fig.~\ref{fig5}. We identified distinct parameter regions where disordered and nematic-ordered solutions are linearly stable. This suggests the existence of a stable nematic state at low $\alpha$ and high density, and a stable disordered state at high $\alpha$ and low density. Near the boundary between these two regions, two instability zones were observed, characterized by the most unstable wavevector being either parallel or perpendicular to the direction of nematic order. 

Previous studies on self-propelled rods predominantly reported the perpendicular instability, which explains the formation of nematic bands observed in the original model \cite{chate2020dry}. In contrast, we identified a parallel instability in this case, similar to findings from analyses of the Vicsek model, where the most unstable wavevector near the ordering transition was found to be parallel to the polar order. Considering the anti-parallel polar bands observed in our microscopic model, we infer that the parallel instability of the nematic order plays a crucial role in their formation. Additionally, the perpendicular instability predicts the emergence of nematic bands, consistent with our simulation results. As $\alpha$ increases, the system undergoes a transition from a homogeneous ordered state to anti-parallel polar bands and then chaotic nematic bands. However, the nematic bands are metastable and may transition to anti-parallel polar bands within finite timescales. To understand this phenomenon, we recall the idea proposed in Ref.~\cite{denk2020pattern}, where particle density not only exhibits pattern formations but also acts as a control parameter for the system's pattern. In fact, the stability diagram reveals a shift in the most unstable wavevector from transverse to longitudinal as the density increases under a fixed $\alpha$. Based on this observation, we propose a possible explanation for the metastability. The continuously bending and deforming nematic bands experience local density fluctuations during their evolution. When these fluctuations exceed the threshold for the aforementioned transition, the nematic band becomes locally unstable in the longitudinal direction. As illustrated in Fig.~\ref{fig4}, this process transforms straight bands into localized anti-parallel polar bands. Over time, these structures undergo transverse diffusion, ultimately leading to a global pattern transition. Moreover, the threshold for this transition, as indicated by the stability diagram, increases with rising $\alpha$. This trend correlates with the progressively extended lifetimes observed in Fig.~\ref{fig3}(b), suggesting that higher values of $\alpha$ make it increasingly difficult for the local density of nematic bands to surpass the threshold required for the transition. 
{Furthermore, with decreasing $\rho$, a contraction and even disappearance of the parallel instability region was observed, alongside an expansion of the perpendicular instability region. This qualitatively reflects the stable nematic band state observed in simulations as the density decreases. 
For the same reason, the transition to metastability with increasing $\alpha$ can also be attributed to a density-induced shift in stability, from an initial perpendicular instability to either a stable nematic solution or parallel instability.
}

\section{\label{sec:level1}Discussion}

In conclusion, we presented a novel self-propelled rods model by introducing frustration through an extended nematic alignment rule, with the separation angle $\alpha$ serving as a key control parameter. Through simulations of the microscopic model, we observed a homogeneous nematic-ordered state at low $\alpha$, while a homogeneous disordered state at high $\alpha$. In the intermediate parameter range, we identified anti-parallel polar bands as well as chaotic nematic bands, which appear as metastable states.

To gain insights into the collective behavior of this model, we studied a corresponding continuum description based on the Boltzmann equation. By expanding the model in angular Fourier modes and truncating the infinite hierarchy, we calculated steady-state homogeneous solutions and conducted a linear stability analysis. The analytical results qualitatively predicted the four states observed in the microscopic model and provided an explanation for the metastability. Specifically, since particle density acts as an pattern control parameter for the system, the local density exceeding a critical threshold within nematic bands triggers longitudinal instabilities, leading to a transition toward anti-parallel polar bands. 

Although the results of the stability diagram exhibit qualitative agreement with the microscopic model, it is important to note the significant differences between the two. 
{In particular, the stable nematic region obtained from the Boltzmann equation is 
significantly narrowed in the microscopic model.}
This discrepancy arises because the Boltzmann approach considers only binary collisions, whereas the frustration in the system primarily originates from many-body interactions.
{Many-body effects in aligning active matter were previously discussed
in terms of polar pattern formation in driven filament systems~\cite{suzuki2015polar}. 
The tunable alignment angle in our model has revealed novel collective patterns 
and further clarified the significance of many-body interactions.

We would like to also mention the difference between the aligning angle 
and the phase lag in the Kuramoto-Vicsek model~\cite{kruk2018self,kruk2020traveling}.
The latter is antagonistic in the sense that the steady-state conditions for a pair,
$\theta_1 - \theta_2 = \alpha$ and $\theta_2 - \theta_1 = \alpha$
are not simultaneously satisified, thereby causing spontaneous rotation.
In contrast,  our alignment rule, $\theta_1 - \theta_2 = \pm \alpha$ 
(for small incoming angles) can be simultaneously accomodated for a pair of particles,
and does not induce chirality.
Additionally, our alignment interaction is soft in the sense that it does not change
the speed of self-propulsion, unlike the excluded volume interaction, 
which slows down particle motion and causes motility-induced phase 
separation~\cite{moran2022particle,moran2022shape}. 
Finally, although the present work considered a two-dimensional system, 
the suppression of homogeneous nematic order by a non-vanishing alignment angle
is expected also in three dimensions.
Experimental studies using cone-shaped colloidal particles with soft interactions 
would be useful to verify our results.
}
{Moreover, mixtures of cones and symmetric rod-like particles also constitute an interesting direction for future research.}
We hope that this study offers insights into the effects of frustration in aligning active matter systems.

The data that support the findings of this article are openly available~\cite{qin2025collective_dataset}.

\bibliography{qin2025collective}

\end{document}